\documentclass[letterpaper]{jpconf}

\usepackage[square,sort&compress]{natbib}
\usepackage[dvips]{graphicx}
\usepackage{dcolumn}
\usepackage{amsmath}
\usepackage{bm}
\usepackage{epsfig}
\usepackage{amssymb}
\begin{document}


\title{Relative importance of crystal field versus bandwidth to the high
pressure spin transition in transition metal monoxides}

\author{Luke Shulenburger,$^{1}$ Sergej Yu Savrasov$^{2}$ and R E
Cohen$^{1,}$\footnote[3]{This paper was presented by R.E.C. as a part of the
invited talk entitled ``Quantum Monte Carlo Simulations of Behavior at
Extreme Conditions''}}

\address{ $^1$ Geophysical Laboratory, Carnegie Institution of
Washington, 5251 Broad Branch Rd, NW, Washington, D. C. 20015, USA}
\address{$^2$ Department of Physics, University of California at Davis, One
Shields Ave, Davis, CA 95616, USA}

\ead{lshulenburger@ciw.edu}

\begin{abstract}
The crystal field splitting and d bandwidth of the $3d$ transition metal
monoxides MnO, FeO, CoO and NiO are analyzed as a function of pressure
within density functional theory.  In all four cases the $3d$ bandwidth is
significantly larger than the crystal field splitting over a wide range of
compressions.  The bandwidth actually increases more as pressure is
increased than the crystal field splitting.  Therefore the role of
increasing bandwidth must be considered in any explanation of a possible
spin collapse that these materials may exhibit under pressure.
\end{abstract}

\section{Introduction}
At low pressures the $3d$ transition metal oxides (TMOs) are magnetic and high
spin.  The existence and nature of a high spin to low spin transition for these materials
at high pressure has been discussed widely in the literature.  The earliest
theoretical studies of the system with density functional theory (DFT) predicted that magnetic
moments on the metal ions switch to a much smaller value  at experimentally
accessible pressures\cite{cohen275mct}.  Subsequent experiments claimed to
confirm this prediction\cite{PhysRevLett.79.5046} only to be contradicted by
later experiments and more refined electronic structure
calculations\cite{PhysRevLett.83.4101,cohen1998mca}.

The question of the mechanism for the spin collapse in $3d$ transition metal
oxides has recently been revisited, with differing conclusions from
dynamical mean field theory\cite{kunes2008cmm} and a model fit to
experimental results\cite{mattila2007mli}.  These papers explain the spin
transition in terms of the crystal field splitting and the $3d$ bandwidth
with Kunes et al. arguing for the crystal field splitting driving the
transition and Mattila et al. suggesting that there is a competition between the
crystal field splitting and the bandwidth.  We address this situation by
calculating the crystal field splitting and $3d$ bandwidth for four TMOs as
a function of pressure within DFT\cite{PhysRev.140.A1133}.

To drive a 
spin collapse via the crystal field splitting, the decreased
exchange energy available to electrons in a high spin configuration must be
offset by the change in the single particle energy levels due to the crystal
field splitting.   The effect of pressure on the exchange energy is small
whereas the crystal field splitting does significantly change the energy of
the d orbitals as the material is compressed.  Typically the crystal field splitting is treated as being
dominated by the electrostatic interaction of the neighboring oxygen atoms
with the metal sites.  However, a much greater effect is the increasing
overlap and subsequent hybridization of the oxygen 2p states with the metal 3d
states\cite{cohen1998mca}.  This change in single particle
energy levels favors the pairing of the d electrons in the lower energy d
orbitals, resulting in a reduced magnetic moment.  

The other explanation for the spin collapse is that as the pressure is
increased, the band energy
sacrificed to break symmetry and form a magnetic state eventually becomes larger than the
energy to be gained by the magnetic interaction.
This competition can be analyzed in light of the extended Stoner theory of
magnetism\cite{anderson1977,PhysRevB.36.8565} which allows these competing
effects to be quantified.  The extended Stoner theory works equally well for
ordered and disordered magnetic states and can predict the optimal magnetic
moment in contrast to the simple Stoner model that only determines if an
instability to magnetism exists.  The essential ingredients to this analysis are
the change in the interaction energy with respect to magnetic moment and the
change in magnetism with respect to the exchange splitting.  Both of these
quantities are driven by a change in the bandwidth near the Fermi level,
which in this case is the $3d$ bandwidth.  In this case an increase in the
bandwidth will raise the energy necessary to form a high spin state,
eventually causing a transition to a low spin state.

\section{Methodology}

We compare these mechanisms by studying an idealized system with a purely cubic structure at zero
temperature in order to isolate the electronic contributions to the
phenomenon (there is significant magnetoelastic coupling in the real monoxides\cite{PhysRevLett.93.215502}).  
Despite the well known failings of DFT withing the local density approximation to capture properties such as
band gap of the TMOs, it is nonetheless useful for gaining a qualitative
intuition into the physics of these materials.  Additionally, nonmagnetic LDA is
the typical starting point for both the LDA+U
\cite{PhysRevB.44.943,PhysRevB.67.153106} and LDA+DMFT\cite{held2006realistic} approaches,
giving a special importance to the details of this solution.  We will also
comment on how the strong correlations present affect these quantities by
examining LDA+U results.

We use a full potential linear muffin tin orbital (LMTO) method, that allows the results of
the DFT calculations to be transformed directly into a tight binding
Hamiltonian with no approximations\cite{PhysRevLett.53.2571}.  In this tight
binding approach, we consider a set of atomic-like wavefunctions centered on the
ions where the each state has energy
\begin{equation}
\epsilon_{\alpha} = \left<\Psi_\alpha\right| \hat{H} \left|\Psi_\alpha\right>
\end{equation}
and states are coupled by hopping matrix elements
\begin{equation}
t_{\alpha,\beta} = \left<\Psi_\alpha\right| \hat{H} \left|\Psi_\beta\right>
\end{equation}
where $\hat{H}$ is an effective Hamiltonian determined from the results of the DFT calculation.

\section{Results and Discussion}
\begin{figure}
\centering
\includegraphics[angle=-90,width=5.5in]{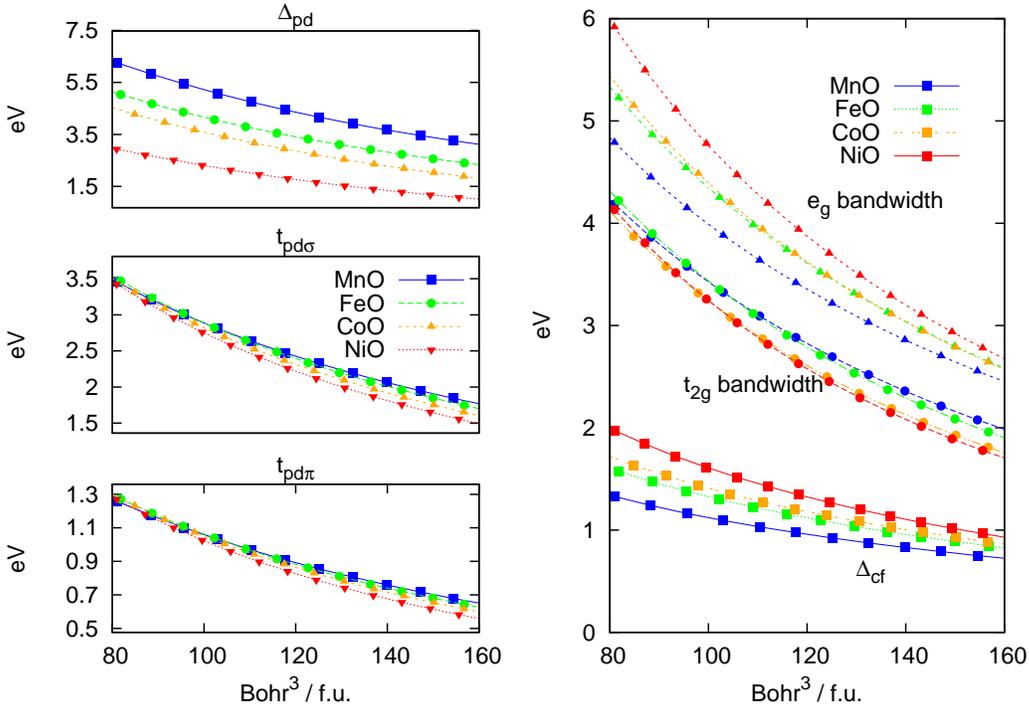}
\caption{Left: Derived tight-binding Hamiltonian elements for the transition metal
oxides versus the functional unit volume (one metal ion and one
oxygen).  $\Delta_{pd}$ is the energy difference
between the $3d\, e_{g}$ orbital and the corresponding oxygen
$2p$ orbital.  Two hopping matrix elements are also shown.   $t_{pd\sigma}$
connects the $t_{2g}$ orbitals with the oxygen $2p$ orbitals parallel to the
line connecting the metal and the oxygen.  Likewise $t_{pd\pi}$ connects the
$t_{2g}$ orbitals with oxygen $2p$ orbitals perpendicular to the line
connecting the metal and the oxygen.  
Right: Derived tight-binding bandwidth and crystal field splitting for  nonmagnetic TMOs via 
LDA calculations as a function of cell volume.  The squares indicate crystal
field splitting, the circles the $t_{2g}$ bandwidth and the triangles the
$e_g$ bandwidth}
\label{nmparams}
\end{figure}

As a first application of this technique, we consider the on site energies of the oxygen
$2p$ orbitals and metal $3d$ state in Fig.\ref{nmparams}.  The plot shows
that as the number of electrons in the problem is increased, the
energy gap between the oxygen and metal $3d$ states decreases and the
hopping between those states also becomes less favorable.  

These calculations also provide estimates of the crystal field splitting and d
bandwidth in the TMOs as a function of pressure, determining the relative size of these terms and how
they behave under pressure.  In this case the bands near the Fermi energy
that are important to the low energy physics of the system are formed by
the hybridization of the metal $3d$ orbitals with the surrounding oxygen $2p$
orbitals.  For the cubic NaCl structure, there are two different symmetries
of d-orbital to be considered.  The first are the two $e_g$
orbitals that point towards the nearest neighbor oxygen
atoms.  These orbitals form bands that are raised in energy versus the isolated d orbital
because of the electrostatic interaction of the electrons on the metal ion
with the negatively charged O ion as well as the hybridization of the p and
d orbitals.  The second class of orbital which is of interest are the three
$t_{2g}$ orbitals that do not point towards the nearest neighbor
oxygens.  These orbitals will form bands which are relatively lower in energy than the $e_g$
orbitals.  The difference in energy between these two sets of orbitals is
conventionally termed crystal field splitting.

To quantify the crystal field splitting and to determine the interactions
that are important to these bands, it is possible to compare the results of
tight binding calculations with ever larger sets of orbitals to the results
of self consistent density functional calculations.  In this case, it is
found that the qualitative features of the bands are preserved by a
Hamiltonian containing only the nearest oxygen $2p$ and metal $3d$ states.
This band would be derived by solving the one dimensional tight binding matrix equation
\begin{equation}
\left( \begin{smallmatrix} \epsilon_d & t_{pd\sigma} \\ t_{pd\sigma} &
\epsilon_p \end{smallmatrix} \right) 
\left(
\begin{smallmatrix} \Psi_d(\vec{k}) \\ \Psi_p(\vec{k}) \end{smallmatrix} 
\right)
= E_{e_g}(\vec{k}) 
\left(
\begin{smallmatrix} \Psi_d(\vec{k}) \\ \Psi_p(\vec{k}) \end{smallmatrix} 
\right),
\label{mateq}
\end{equation}
where the basis for the wavefunction is the proper symmetry $d$ orbital on
the metal ion and the $p$ orbital on the
nearest neighbor oxygen ions.  In this case the crystal momentum $\vec{k}$
is given along the direction connecting the metal ion and one of the nearest
neighbor oxygens, which is appropriate because the band obtains both its
minimum and maximum energies in this direction.  Solving this equation for
the lowest energy state gives an energy for the $e_g$ band
\begin{equation}
E_{e_g}(\vec{k}) = \frac{\epsilon_{d} + \epsilon_{p}}{2} +
\sqrt{\left(\frac{\epsilon_{d} -
\epsilon_{p}}{2}\right)^2 + 4 t_{pd\sigma}^2
\sin\left(\frac{\vec{k}\cdot\vec{a}}{2}\right)},
\label{nnen}
\end{equation}
where $\vec{a}$ is a primitive translation vector in the same direction as
$\vec{k}$.

\begin{figure}
\centering
\includegraphics[angle=0,width=3.5in]{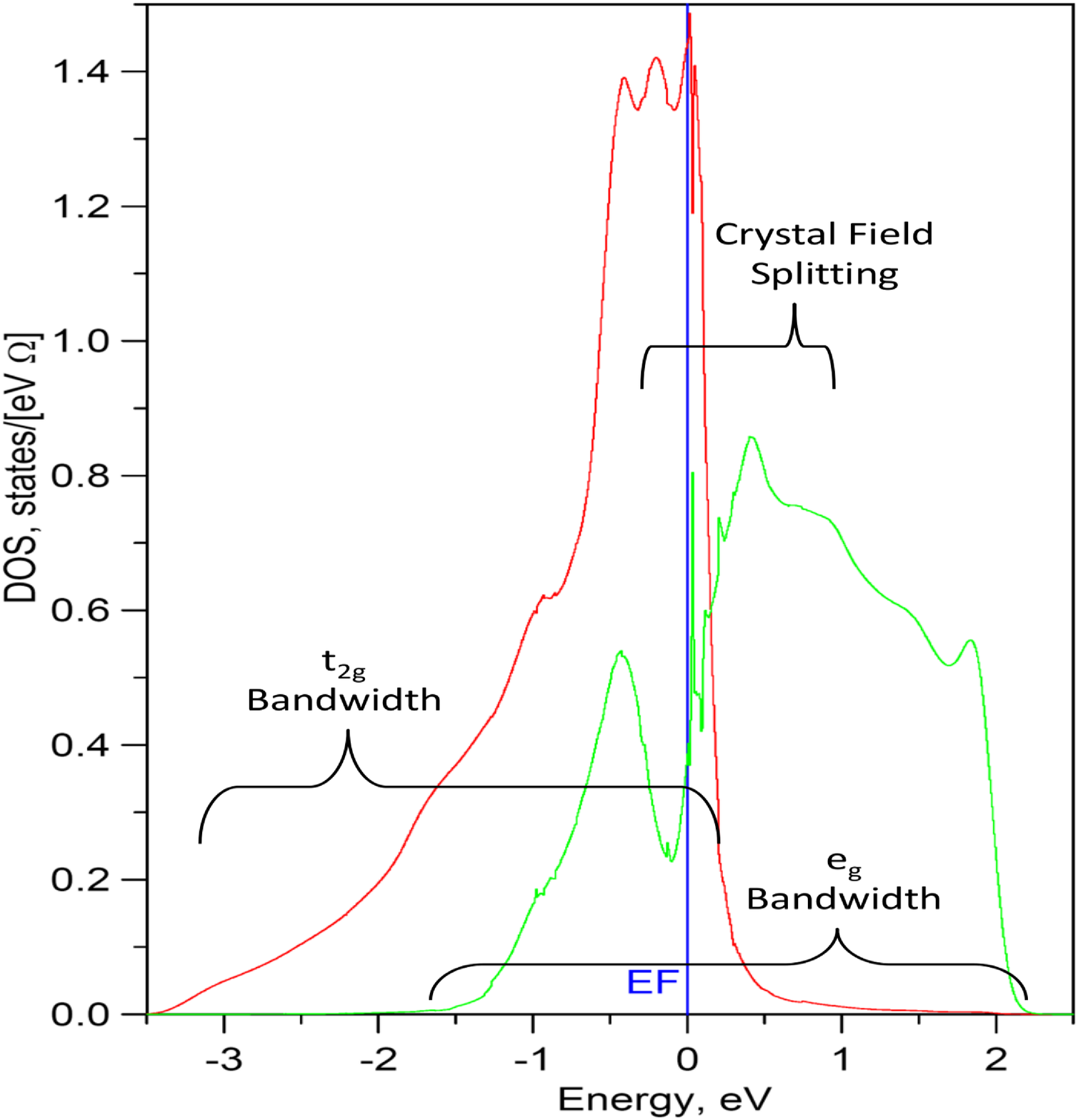}
\caption{Partial density of states of FeO with a lattice constant of 4.0228
Angstrom from DFT LDA calculations.  The
total density of states is projected onto atomic d orbitals centered on the
Fe atom.  The bandwidths of the $e_g$ and $t_{2g}$ bands are indicated, as
well as the crystal field splitting which is the distance between the
centers of the two bands.  All energies are measured in eV relative to the
Fermi energy.}
\label{dosfig}
\end{figure}

For quantitative agreement in the width and location of the bands, it is
necessary to include next nearest neighbor hopping between the metal ion and the
fcc lattice of metal ions surrounding it.  This can be done by replacing the
onsite energy, $\epsilon_d$, of the d orbital with a $\vec{k}$ dependent
band taking into account the hybridization with the next nearest neighbor metal ions.
 Following the work of Harrison\cite{harrison-book} and solving a similar but larger matrix
equation as in Eq.\,\ref{mateq}, this gives the following expression for the
energy of the $d$ band:
\begin{equation}
E_{d}(\vec{k}) = \epsilon_{d} + t_{dd\pi} + 3
t_{dd\delta} + \left(\frac{t_{dd\sigma}}{2}
+ \frac{3 t_{dd\pi}}{2} + \frac{3 t_{dd\delta}}{2} \right) \cos
\left(\frac{\sqrt{2} \vec{k}\cdot\vec{a}}{2}\right)
\label{tb2}
\end{equation}
Substituting this $k$ dependent energy of the $d$ band into
Eq.\,\ref{nnen}, the full energy dependence of the $e_g$ band including
next nearest neighbor interactions is 
\begin{equation}
E_{e_g}(\vec{k}) = \frac{E_{d}(\vec{k}) + \epsilon_{p}}{2} +
\sqrt{\left(\frac{E_{d}(\vec{k}) -
\epsilon_{p}}{2}\right)^2 + 4 t_{pd\sigma}^2 \sin\left(\frac{\vec{k}\cdot\vec{a}}{2}\right)}
\label{tb1}
\end{equation}

A similar expression can be obtained for the $t_{2g}$ band, first
substituting $t_{pd\pi}$ for $t_{pd\sigma}$ in Eq.~\ref{tb1} and using the appropriate matrix
elements for the second nearest neighbor $d-d$ interaction in Eq.~\ref{tb2}.  Using
these expressions in Eqs.~\ref{tb2} and \ref{tb1}, the bandwidth is the
difference between the extremes of Eq.~\ref{tb1} and the crystal field
splitting is the difference between the center of mass of the $e_g$ and $t_{2g}$ bands.
With this prescription, the bandwidth and crystal field splitting of the
transition metal oxides as a function of lattice constant is shown on the
right side of Fig.~\ref{nmparams}

In order to check the validity of these tight binding results, we calculate the bandwidth
and crystal field splitting of FeO using both the tight binding
approximation and the partial density of states.  For the partial density of
states, the $d$ bandwidth is simply the spread of the density of states
projected onto the $3d$ orbitals with the same symmetry as the band in question.  The
crystal field splitting is determined by the distance between the center of
mass of the two independent $3d$ bands.  These quantities are illustrated
for the partial density of states of FeO as shown in Fig.~\ref{dosfig}. 

\begin{figure}
\centering
\includegraphics[angle=-90,width=5.5in]{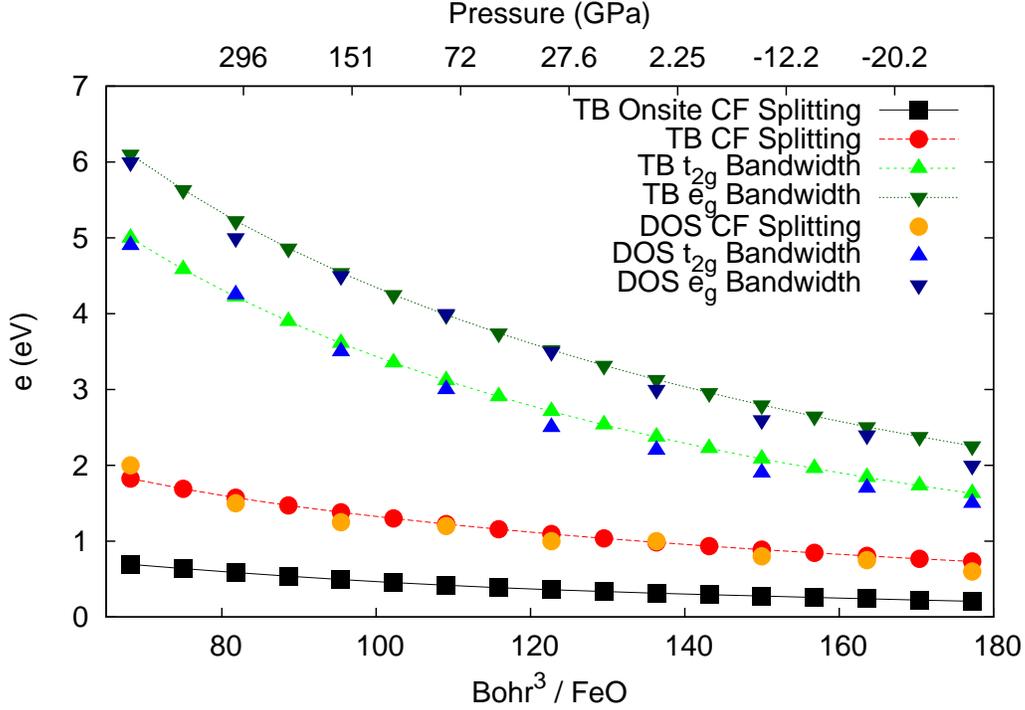}
\caption{Crystal field splitting and $3d$ bandwidth of FeO
from tight binding and the density of states.  The black squares are the
crystal field splitting from the on site terms in the tight binding.  Points
connected with lines are from the tight binding expressions.  All quantities
are plotted as a function of the volume of the cell containing one FeO pair.  
The pressures indicated on the top axis are from a Vinet equation of state fit to 
the antiferromagnetic LDA+U energies.}
\label{tb-vs-dos}
\end{figure}

The agreement between these two methods is excellent as shown in
Fig.~\ref{tb-vs-dos}.  This figure also shows that the crystal field
splitting is not simply due to a change in the energy of the various $3d$
orbitals on the metal ion.  This result is somewhat surprising given the seminal work of
Mattheiss who showed from symmetry arguments that the crystal field
splitting at the gamma point is due entirely to hopping between nearest
neighbor metal ions with no effect due to the
oxygen\cite{PhysRevB.5.290,PhysRevB.5.306}.  This seeming inconsistency is
resolved by noting that the full crystal field splitting is an average over
the full Brillouin zone rather than being defined by the behavior at the
gamma point.  Indeed, the Mattheiss result is not incompatible with our tight
binding framework.  In tight binding calculations, the hopping matrix element between the
O $2p$ and Fe $3d$ states can be set to zero independently with the
result that the crystal field splitting at gamma is unchanged, but the full
crystal field splitting is changed.

\begin{figure}
\centering
\includegraphics[angle=-90,width=5.5in]{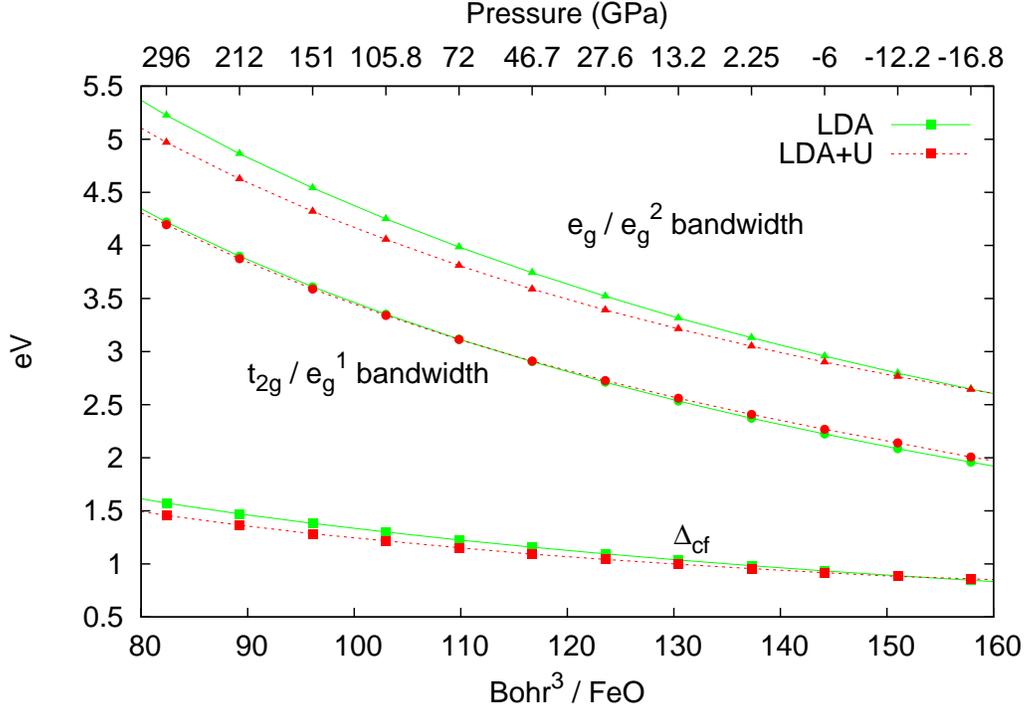}
\caption{Comparison of bandwidth and crystal field splitting obtained via 
nonmagnetic LDA and antiferromagnetic LDA+U calculations as a function of
compression.  The squares indicate crystal
field splitting, the circles the $t_{2g}$ bandwidth and the triangles the
$e_g$ bandwidth.  The LDA results are plotted in green and connected by
solid lines and the LDA+U results are plotted in red and connected by dotted
lines.  The pressures are from a Vinet equation fit to the LDA+U energies.}
\label{comp-lda-ldau}
\end{figure}

These calculations have shown that the absolute value of the $3d$ bandwidth
is much greater than the crystal field splitting.  Additionally, the change
of this quantity with pressure is much greater than that of the the crystal
field splitting.  These results strongly suggest that the spin transition is
not entirely due to the change of crystal field splitting with pressure,
owing at least in part to the increase in the $3d$ bandwidth.  

To check the robustness of this result, the crystal field splitting and $3d$ bandwidth
were also calculated using LDA+U with the tight binding approach.  In the
antiferromagnetic ground state, the electronic
symmetry is lowered from cubic to rhombohedral thus the crystal field
picture is slightly changed.   However, the 
difference between the $e_g^1$ state and the $e_g^2$ state for the minority
spin of FeO at ambient
pressure gives a crystal field splitting of $0.95$ eV, and bandwidths of
$3.05$ and $2.408$ eV respectively with a dependence on pressure which is
very similar to that of the LDA results.  This result is robust with respect
to pressure, as shown for FeO in Fig.~\ref{comp-lda-ldau}, showing that addition of
the magnetism and strong interactions does not greatly change the nature of
bonding for the $3d$ electrons.

\section{Conclusion}
We have shown that the 3d bandwidth in MnO, FeO, CoO and
NiO is greater than the crystal field splitting and increases with
pressure.  This result is robust to the choice of LDA or LDA+U correlations
for the $3d$ electrons.  Explanations of the spin collapse in these
materials should account for this increasing bandwidth as well as the change
in crystal field splitting.

\ack
We wish to thank I. Mazin, H-K. Mao, S. Gramsch, and P. Ganesh for insightful
discussions.  L.S. and R.E.C acknowledge support in the form
of NSF grants TG-MCA07S016, EAR-0738061 and EAR-0530282.  The LMTO code used
for the calculations was developed by Sergej Savrasov with support from NSF
grant DMR-0606498.

\bibliography{tmo}

\end{document}